\documentclass[pra, twocolumn, 
 showpacs, floatfix, superscriptaddress]{revtex4}
\usepackage{amsmath}
\usepackage{amssymb}
\usepackage{latexsym}
\usepackage{bbm}
\usepackage{bm}
\usepackage{amsthm}
\usepackage{graphicx}

\theoremstyle{plain}

\begin{document}
\title{Topological Quantum Computing with $p$-Wave Superfluid
Vortices}
\author{Tetsuo Ohmi}
\affiliation{Research Center for Quantum Computing,\\Interdisciplinary Graduate School of Science
and Engineering, Kinki University, Higashi-Osaka 577-8502, Japan}
\email{ohmi@math.kindai.ac.jp}
\author{Mikio Nakahara}
\affiliation{Research Center for Quantum Computing,\\Interdisciplinary Graduate School of Science
and Engineering, Kinki University, Higashi-Osaka 577-8502, Japan}
\email{ohmi@math.kindai.ac.jp}
\affiliation{Department of Physics, Kinki University, Higashi-Osaka
577-8502, Japan}
\email{nakahara@math.kindai.ac.jp}

\begin{abstract}
It is shown that Majorana fermions trapped in three vortices in
a $p$-wave superfluid form a qubit in a topological quantum computing (TQC).
Several similar ideas have already been proposed: Ivanov 
[Phys. Rev. Lett. {\bf 86}, 268 (2001)] and
Zhang {\it et al.} [Phys. Rev. Lett. {\bf 99}, 220502 (2007)]
have proposed schemes in which a qubit is implemented
with two and four Majorana fermions, respectively, where
a qubit operation is performed by exchanging the 
positions of Majorana fermions. The set of
gates thus obtained is a discrete subset of the
relevant unitary group. We propose, in this paper, a new scheme,
where three Majorana fermions form a qubit. We show that continuous
1-qubit gate operations are possible by exchanging the 
positions of Majorana fermions complemented with dynamical phase
change. 2-qubit
gates are realized through the use of the coupling between Majorana
fermions of different qubits.
\end{abstract}
%
%

\maketitle

\section{Introduction}

Ivanov first pointed out that a pair of Majorana fermions can be
used to implement a qubit and proposed gate operations on it \cite{I01}.
He has also demonstrated that a braiding of Majorana fermions
leads to entanglement of two qubits. Later, Zhang {\it et al.}
proposed to use four Majorana fermions to implement a qubit \cite{ZTD07}.
They further proposed to use a flying qubit to entangle 
two qubits thus implemented. It should be noted, however,
that a braiding is a discrete operation and it is impossible 
to implement an arbitrary one-qubit gate with a braiding.
Moreover, it should be also pointed out that entangling operation 
using a flying qubit does not work in practice, since the Majorana
fermion does not couple with density fluctuation as shown in \cite{Chung}.
It is the purpose of this paper to show that continuous gate
operations are possible if a qubit is implemented with 
three Majorana fermions. We use two Majorana fermions, similarly to
Ivanov's proposal, to implement a qubit and an additional Majorana
fermion for continuous control of the qubit state. Similarly continuous 2-qubit
gates can be implemented by making use of the coupling between Majorana
fermions which belong to different qubits.

Let us consider a $p$-wave superfluid with the order parameter
$p_x+i p_y$. A vortex in the superfluid supports
a bound state in the quasiparticle spectrum,
whose bound state energy is exactly at
the center of the band gap. The bound state is invariant
under charge conjugation and called the Majorana mode,
which will be called the Majorana fermion hereafter \cite{GR07}.
It has been shown by Mizushima, Ichioka and Machida that this
zero-energy state is energetically well separated from the
other bound states (Caroli-de Gennes-Matericon states)
in the strong coupling limit, in which the energy gap $\Delta$ is on
the same order as the Fermi energy $E_{\rm F}$ \cite{MIM08}. 
Topological quantum computing employs Majorana fermions
in such strongly coupled systems \cite{RMP08}.

Let us consider a two-Majorana fermion system, first.
The Hamiltonian of this system is given by
\begin{equation}
H = i J_{12} \gamma_1 \gamma_2,
\end{equation}
where $J_{12}$ is the coupling constant between two Majorana fermions
and $\gamma_i$ stands for the Majorana operator associated with
the $i$th vortex.
They satisfy the anticommutation relation
\begin{equation}
\{ \gamma_i, \gamma_j\} = 2 \delta_{ij}.
\end{equation}
We now introduce another set of operators $\alpha$ and $\alpha^{\dagger}$
\begin{equation}
\alpha= \frac{1}{2}(\gamma_1+i \gamma_2),
\ \alpha^{\dagger}= \frac{1}{2}(\gamma_1-i \gamma_2),
\end{equation}
which satisfy the fermion anticommutation relation
\begin{equation}
\{ \alpha, \alpha \} = \{ \alpha^{\dagger}, \alpha^{\dagger} \}=0,
\quad
\{ \alpha, \alpha^{\dagger} \} = 1.
\end{equation}
The Hamiltonian is then rewritten, in terms of the new operatos, as
\begin{equation}
H= \omega \left(2\alpha^{\dagger} \alpha-1\right),\quad \omega = J_{12}.
\label{eq:hamalpha}
\end{equation}

It is shown that the Bogoliubov wave functions $u(\bm{r})$
and $v(\bm{r})$ satisfy the relation $u(\bm{r}) = v^*(\bm{r})$
for a zero-energy mode
and hence the Majorana operator is expressed as $\gamma_i 
= c_i + c_i^{\dagger}$, where $c_i = \int d^2 \bm{r} u_i(\bm{r})^*
\psi(\bm{r})$. Here $\psi(\bm{r})$ is the field operator of the
particles in $p$-wave superfluid state and $u_i(\bm{r})$ is
the Bogoliubov wave function of the zero-energy state trapped
in the $i$th vortex. Let $|0 )_i$,
defined by $c_i |0 )_i=0$, denote the state in which
the $i$th vortex has no zero-energy particle, while $c^{\dagger}_i|0
)_i =|1 )_i$
denote the state with a zero-energy particle at the $i$th
vortex.

The ground state energy of this Hamiltonian (\ref{eq:hamalpha})
is $-\omega$, which has two eigenvectors 
\begin{equation}
\alpha |0 )_1 | 0 )_2,\quad \alpha \alpha^{\dagger}
|0 )_1 | 0 )_2,
\end{equation}
where the first eigenvector has odd fermion number while the second one
has even fermion number. The excited state energy is $+\omega$, 
which is doubly degenerate with the energy eigenstates
\begin{equation}
\alpha^{\dagger} |0 )_1 | 0 )_2,\quad \alpha^{\dagger} \alpha
|0 )_1 | 0 )_2,
\end{equation}
where the first eigenvector again has odd fermion number while the second one
has even fermion number. 

We note that application of $\alpha^{\dagger}$ on the ground states
changes the parity of the fermion number as
\begin{equation}
\begin{array}{c}
\alpha|0 )_1|0 )_2 \to \alpha^{\dagger} \alpha|0 )_1|0 )_2 
\\
\\
\alpha \alpha^{\dagger} |0 )_1|0 )_2 \to 
\alpha^{\dagger} |0 )_1|0 )_2.
\end{array}
\end{equation}

\section{Three-Majorana Fermion Model}

Suppose there are three vortices, each of which supports
a Majorana fermion in the limit of infinite separations
among the vortices. The Hamiltonian describing
the coupled Majorana fermions is given by
\begin{equation}
H = i J_{12} \gamma_1 \gamma_2+ i J_{23} \gamma_2 \gamma_3 
+ i J_{31} \gamma_3 \gamma_1,
\label{eq:hamiltonian}
\end{equation}
where $J_{ij} \in \mathbb{R}$ is the coupling strength
between the $i$th and the $j$th Majorana fermions.
It turns out to be convenient to parametrize three coupling
constants by the polar angles $\theta$ and $\phi$ as
\begin{equation}
(J_{23}, J_{31}, J_{12}) = J(\sin \theta \cos \phi,
\sin \theta \sin \phi, \cos \theta).
\end{equation}

The Hamiltonian is diagonalized by introducing the creation
and the destruction operators
\begin{equation}
\begin{array}{l}
\alpha^{\dagger} =
\frac{1}{2} \left[ (\cos \theta \cos \phi+ i \sin \phi)
\gamma_1\right. \vspace{.2cm}\\
\qquad \qquad \left. + (\cos \theta \sin \phi -i \cos \phi ) \gamma_2 - 
\sin \theta \gamma_3\right]
\vspace{.2cm}\\
\alpha =
\frac{1}{2} \left[ (\cos \theta \cos \phi- i \sin \phi)
\gamma_1 \right. \qquad \qquad \\
\qquad \qquad \left.
 + ( \cos \theta \sin \phi+i \cos \phi) \gamma_2 - 
\sin \theta \gamma_3\right]
\end{array}
\label{eq:cd}
\end{equation}
as
\begin{equation}
H = \omega \left( 2 \alpha^{\dagger} \alpha -1 \right),
\end{equation}
where $\omega = J$. It should be noted that there exists
a Majorana fermion operator
\begin{equation}
\beta = \sin \theta \cos \phi \gamma_1 + \sin \theta \sin \phi
\gamma_2 + \cos \theta \gamma_3,
\end{equation}
which is orthogonal to $\alpha$ and $\alpha^{\dagger}$.
It is easy to verify 
these fermionic operators satisfy the anticommutation relations
\begin{equation}
\{\alpha, \alpha^{\dagger}\} = 1,
\ \{\alpha, \beta\}= \{\alpha^{\dagger}, \beta \} = 0.
\end{equation}
It follows from the above anticommutation relations that
$\beta$ commutes with $H$ and, hence, $\beta$ represents
the zero-energy Majorana fermion. Mizushima and Machida 
analyzed the lowest energy eigenvalues by solving the Bogoliubov-de Gennes 
equation numerically and obtained the same results \cite{MM09}.

The operators $\alpha, \alpha^{\dagger}$ and $\beta$ take the simpler
forms
\begin{equation}
\alpha = \frac{e^{-i \phi}}{2}(\gamma_1 + i \gamma_2),
\alpha^{\dagger} = \frac{e^{i \phi}}{2}(\gamma_1 - i \gamma_2),
\beta = \gamma_3
\end{equation}
in the limit $J_{12} \gg J_{23}, J_{31}$, which corresponds to the case
in which vortex 3 is isolated from vortices 1 and 2.
We also have
\begin{equation}
\tan \phi = \frac{J_{31}}{J_{23}}.
\end{equation}

Now let us analyze the energy eigenstates of the Hamiltonian
(\ref{eq:hamiltonian}) in the above limit.
The ground state with the energy $-\omega$ is four-fold
degenerate. Ground states with odd number
of Majorana fermions are two-fold degenerate,
\begin{eqnarray}
\alpha |0 )_1 |0 )_2 |0 )_3
&=& \frac{e^{-i \phi}}{2}(\gamma_1 + i \gamma_2)|0 )_1
|0 )_2 |0 )_3 \nonumber\\
&=& \frac{e^{-i \phi}}{2}\left( |1 )_1 |0 )_2 |0 )_3 
+i |0 )_1 |1 )_2 |0 )_3 \right)\nonumber\\
& &\\
\alpha \alpha^{\dagger} \beta |0 )_1 |0 )_2 |0 )_3
&=& \frac{1}{2}(\gamma_3-  i \gamma_1 \gamma_2 \gamma_3)|0 )_1
|0 )_2 |0 )_3 \nonumber\\
&=& \frac{1}{2}\left( |0 )_1 |0 )_2 |1 )_3 
-i |1 )_1 |1 )_2 |1 )_3 \right).\nonumber\\
& &
\end{eqnarray}
Similarly, ground states with even number
of Majorana fermions are two-fold degenerate with the eigenstates 
\begin{eqnarray}
\alpha  \alpha^{\dagger}|0 )_1 |0 )_2 |0 )_3
&=& \frac{1}{2}(1- i \gamma_1 \gamma_2)|0 )_1
|0 )_2 |0 )_3 \nonumber\\
&=& \frac{1}{2}\left( |0 )_1 |0 )_2 |0 )_3 
-i |1 )_1 |1 )_2 |0 )_3 \right)\nonumber\\
& & \\
\alpha \beta  |0 )_1 |0 )_2 |0 )_3
&=& \frac{e^{-i \phi}}{2} (\gamma_1+i \gamma_2)\gamma_3 |0 )_1
|0 )_2 |0 )_3 \nonumber\\
&=& \frac{e^{-i \phi}}{2}\left( |1 )_1 |0 )_2 |1 )_3 
+i |0 )_1 |1 )_2 |1 )_3 \right).\nonumber\\
& &
\end{eqnarray}

The excited state with the energy $\omega$ is also four-fold degenerate;
states with odd fermion number are
\begin{eqnarray}
\alpha^{\dagger}|0 )_1 |0 )_2 |0 )_3
&=& \frac{e^{i \phi}}{2}(\gamma_1- i  \gamma_2)|0 )_1
|0 )_2 |0 )_3 \nonumber\\
&=& \frac{e^{i \phi}}{2}\left( |1 )_1 |0 )_2 |0 )_3 
-i |0 )_1 |1 )_2 |0 )_3 \right)\nonumber\\
& &\\
\alpha^{\dagger} \alpha \beta  |0 )_1 |0 )_2 |0 )_3
&=& \frac{1}{2} (\gamma_3+i\gamma_1 \gamma_2\gamma_3)|0 )_1
|0 )_2 |0 )_3 \nonumber\\
&=& \frac{1}{2}\left( |0 )_1 |0 )_2 |1 )_3 
+i |1 )_1 |1 )_2 |1 )_3 \right),\nonumber\\
& &
\end{eqnarray}
while those with even fermion numbers are
\begin{eqnarray}
\alpha^{\dagger}\alpha |0 )_1 |0 )_2 |0 )_3
&=& \frac{1}{2}(1+ i \gamma_1 \gamma_2)|0 )_1
|0 )_2 |0 )_3 \nonumber\\
&=& \frac{1}{2}\left( |0 )_1 |0 )_2 |0 )_3 
+i |1 )_1 |1 )_2 |0 )_3 \right)\nonumber\\
& &\\
\alpha^{\dagger} \beta  |0 )_1 |0 )_2 |0 )_3
&=& \frac{e^{i \phi}}{2} (\gamma_1-i \gamma_2)\gamma_3|0 )_1
|0 )_2 |0 )_3 \nonumber\\
&=& \frac{e^{i \phi}}{2}\left( |1 )_1 |0 )_2 |1 )_3 
-i |0 )_1 |1 )_2 |1 )_3 \right).\nonumber\\
& &
\end{eqnarray}

Transitions among the ground states and the excited states
could be performed by Rabi oscillation through modulation in $J_{23}$
or $J_{31}$. Suppose the interactions
\begin{equation}
\begin{array}{c}
i \delta J_{23} \gamma_2 \gamma_3 \cos 2 \omega t
= -\delta J_{23} (\alpha^{\dagger} e^{-i \phi}-\alpha e^{i \phi}) \beta 
\cos 2 \omega t
\vspace{0.2cm}\\
i \delta J_{31} \gamma_3 \gamma_1 \cos 2 \omega t
= -i \delta J_{31} (\alpha^{\dagger} e^{-i \phi}+ \alpha e^{i \phi}) \beta 
\cos 2 \omega t
\end{array}
\end{equation}
are introduced in the Hamiltonian. Then the following Rabi oscillations
take place between the four sets of states;
\begin{equation}
\begin{array}{ccc}
\mbox{ground state}& &\mbox{excited state}\vspace{.2cm}\\
\alpha |0 )_1 |0 )_2 |0 )_3&\leftrightarrow&
 \alpha^{\dagger} \alpha \beta |0 )_1 |0 )_2 |0 )_3
\vspace{.2cm}\\
\alpha \alpha^{\dagger} \beta 
|0 )_1 |0 )_2 |0 )_3&\leftrightarrow&
\alpha^{\dagger}  |0 )_1 |0 )_2 |0 )_3
\vspace{.2cm}\\
\alpha \alpha^{\dagger} |0 )_1 |0 )_2 |0 )_3&\leftrightarrow&
  \alpha^{\dagger} \beta |0 )_1 |0 )_2 |0 )_3
\vspace{.2cm}\\
\alpha \beta |0 )_1 |0 )_2 |0 )_3&\leftrightarrow&
 \alpha^{\dagger} \alpha |0 )_1 |0 )_2 |0 )_3.
\vspace{.2cm}\\
\end{array}
\end{equation}
Note that the Rabi oscillations preserve the parity of the fermion
number. It is possible to implement a continuous series of
quantum gate operations by making use of the above Rabi oscillations.
However, this may cause qubit operation error since the system is
under external field, which possibly contains noise.
It is certainly desirable to perform
qubit operations without errors by exchanging the vortex
positions as was proposed by Ivanov \cite{I01} and Zhang {\it et al.}
\cite{ZTD07}.

Now we turn to our main result, in which continuous qubit
operations are implemented by introducing dynamical phases in TQC.

\section{One-Qubit Gates}

Let us first consider the odd fermion number sector with 
the initial state
\begin{eqnarray*}
\alpha |0 )_1 |0 )_2 |0 )_3
&=& \frac{e^{-i \phi}}{2} (\gamma_1 + i \gamma_2)
|0 )_1 |0 )_2 |0 )_3\\
&=& \frac{e^{-i \phi}}{2} (|1 )_1 |0 )_2 |0 )_3
+i |0 )_1 |1 )_2 |0 )_3).
\end{eqnarray*}
We assume the vortices at 1 and 2 are also remotely separated 
initially so that all the coupling strengths are small. We still
impose the condition $J_{12} \gg J_{23}, J_{31}$ even in this case.
Then the dynamical phase changes for the ground states and the excited
states are almost identical since $\omega = J$ is negligibly small.
Now we outline how to implement a unitary gate with continuous parameters
in several steps as shown in Fig.~1.
\begin{figure*}
\begin{center}
\includegraphics[width=10cm]{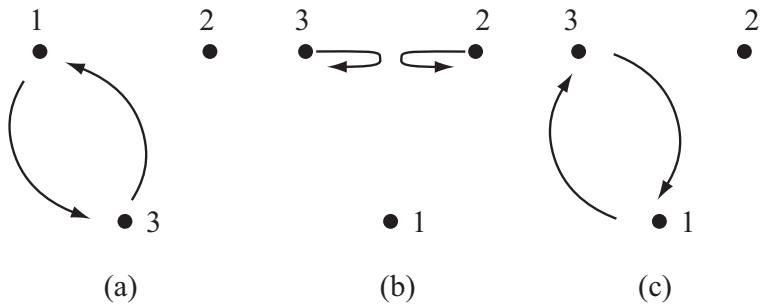}
\end{center}
\caption{Implementation of a one-qubit gate. Numbers 1, 2 and 3
show the positions of vortices.
(a) Vortices at positions 1 and 3 are
exchanged (STEP 1). (b) Vortices at 1 and 2 are put close to each other
so that they acquire the dynamical phase (STEP 2). (c) Vortices at 1 and 3
are exchanged again so that the vortices take their initial 
configuration (STEP 3).}
\end{figure*}
\begin{enumerate}
\item[STEP 1] Suppose vortices at positions 3 and 1 are exchanged 
in the counterclockwise sense, as shown in Fig.~1~(a),
so that the 
Majorana operators are transformed as $\gamma_3 \to \gamma_1$
and $\gamma_1 \to - \gamma_3$. Under this transformation,
the operator $\alpha$ transforms as
\begin{eqnarray*}
\alpha=
\frac{e^{-i\phi}}{2}(\gamma_1 + i \gamma_2) &\to& 
\frac{e^{-i\phi}}{2}(-\gamma_3 + i \gamma_2)\\
&=& -\frac{e^{-i\phi}}{2}(\alpha \alpha^{\dagger}\beta  + \alpha^{\dagger}
\alpha \beta)\\
& &+ \frac{e^{-i\phi}}{2}(\alpha e^{i \phi} - \alpha^{\dagger} e^{-i \phi})
\end{eqnarray*}
Transformations of the
operators $\alpha^{\dagger}, \alpha \alpha^{\dagger} \beta $
 and $ \alpha^{\dagger} \alpha \beta$ under
this exchange are also obtained and summarized as
\begin{equation}
\left( \begin{array}{c}
\alpha\\
\alpha^{\dagger} \alpha \beta \\
\alpha \alpha^{\dagger} \beta \\
\alpha^{\dagger}
\end{array} \right)
\to m_{31}
\left( \begin{array}{c}
\alpha\\
\alpha^{\dagger} \alpha \beta \\
\alpha \alpha^{\dagger} \beta \\
\alpha^{\dagger}
\end{array} \right)
\end{equation}
where
\begin{equation}
m_{31} = \frac{1}{2}
\left( \begin{array}{cccc}
1&-e^{-i \phi}& -e^{-i \phi} &-e^{-2i \phi}\\
e^{i \phi}& 1 &-1&e^{-i \phi}\\
e^{i \phi}& -1&1&e^{-i \phi}\\
-e^{2i \phi}& -e^{i \phi}& -e^{i \phi}&1
\end{array} \right)
\label{eq:m31}
\end{equation}
\item[STEP 2] Vortices at 1 and 2 are put close to each other,
as shown in Fig.~1~(b), so that
$J_{12}$ is appreciably large. Now both the ground state and the
excited states acquire nontrivial phases. The transformation matrix
is 
\begin{equation}
m_z = \left( \begin{array}{cccc}
e^{-i \eta}&0&0&0\\
0& e^{i \eta}&0&0\\
0&0&e^{-i \eta}&0\\
0&0&0&e^{i \eta}
\end{array} \right).
\end{equation}
\item[STEP 3] Subsequently, vortices at 3 and 1 are exchanged 
in clockwise sense as shown in Fig.~1~(c), which
introduces $m_{31}^{-1}$.
\end{enumerate}

The above three steps result in a transformation matrix
\begin{eqnarray}
\lefteqn{m_{31}^{-1} m_z m_{31}}\nonumber\\
&=& \left( \begin{array}{cccc}
\cos \eta & -i e^{-i \phi} \sin \eta& 0&0\\
- i e^{i \phi} \sin \eta & \cos \eta& 0&0\\
0&0& \cos \eta& i e^{-i \phi} \sin \eta\\
0&0& i e^{i \phi}\sin \eta&  \cos \eta
\end{array} \right).\nonumber\\
& &
\label{eq:gate}
\end{eqnarray}
This result shows that the qubit basis vectors $|0 \rangle = \alpha
|0 )_1 |0 )_2 |0 )_3$ and $|1 \rangle
=  \alpha^{\dagger} \alpha \beta |0 )_1 |0 )_2 |0 )_3$
are continuously transformed. This statement remains true if another
set of the qubit basis vectors, $|0 \rangle =\alpha \alpha^{\dagger} \beta
|0 )_1 |0 )_2 |0 )_3$ and $|1 \rangle
= \alpha^{\dagger} |0 )_1 |0 )_2 |0 )_3$, are chosen.

It is instructive to implement the Hadamard gate 
$$
U_{\rm H} = \frac{1}{\sqrt{2}} \left(\begin{array}{cc}
1&1\\
1&-1
\end{array} \right)
$$
with our scheme. We use $|0 \rangle = \alpha
|0 )_1 |0 )_2 |0 )_3$ and $|1 \rangle
=  \alpha^{\dagger} \alpha \beta |0 )_1 |0 )_2 |0 )_3$
as the qubit basis. Then the upper-left block of the matrix (\ref{eq:gate})
has relevance. Let us write
\begin{equation}
M(\eta, \phi) = \left( \begin{array}{cc}
\cos \eta & -i e^{-i \phi} \sin \eta\\
- i e^{i \phi} \sin \eta & \cos \eta
\end{array}
\right).
\label{qtc:eq:mgate}
\end{equation}
Then we easily verify the product $M(\frac{\pi}{4}, -\frac{\pi}{2})
M(\frac{\pi}{2},0)$ implements the Hadamard gate up to an overall phase.

Qubit operations are also possible by exchanging vortices at 2 and 3,
instead of vortices at 1 and 2. It is also easy to verify that
a similar qubit construction and qubit operations are 
possible if the qubit basis states are made of even fermion number states.
The sequence of operations given in Fig.~1, in this case, results in the 
matrix (\ref{eq:gate}), although $m_{31}$ takes a different
form from the odd fermion case (\ref{eq:m31}).

It has been shown so far that a continuous family of 1-qubit operations
can be implemented by adding a third Majorana fermion to a pair of
Majorana fermions.

\section{Two-Qubit Gates}

Finally, we show that our qubits satisfy the universality
criterion by demonstrating that two-qubit gates can be implemented
within the current proposal. We first note that  
the third Majorana fermion is required only to implement single-qubit gates
and plays no role if it is far remote from the 
first and the second Majorana fermions. Let us first consider the
braiding proposed in \cite{I01}. Let $\gamma_1$ and $\gamma_2$
($\gamma'_1$ and $\gamma'_2$) be 
the Majorana fermion operators associated with qubit 1 (2),
where an index associated with the second qubit is denoted with a prime.
Let 
the initial state of qubits 1 and 2 be $\alpha \alpha' |0 \rangle$, where
$$
\alpha = \frac{1}{2}(\gamma_1 + i \gamma_2),
\ \alpha'= \frac{1}{2}(\gamma'_1 + i \gamma'_2)
$$
and we write $|0 )_{1} | 0)_{2} |0 )_{1'}, 
|0 )_{2'}$ as
$|0 )$ to simplify the notation. 
Ivanov [\cite{I01}] attempted to create an entangled state $\frac{1}{\sqrt{2}}
(|0 \rangle |0 \rangle + |1 \rangle |1 \rangle)$ by braiding
 of Majorana fermions. 
Let us exchange Majorana fermions
1 and 1' in the counterclockwise sense. The state then transforms as
$$
\alpha \alpha' | 0 ) \to \frac{1}{2}(\alpha \alpha'
+ \alpha^{\dagger} \alpha'^{\dagger} - \alpha \alpha^{\dagger}
\alpha' \alpha'^{\dagger} + \alpha^{\dagger} \alpha
\alpha'^{\dagger} \alpha')| 0 ),
$$
which is certainly an entangled state. However, this state is different
from the state
\begin{equation}
\frac{1}{\sqrt{2}}(|0 \rangle |0 \rangle + |1 \rangle |1 \rangle)=
\frac{1}{\sqrt{2}}(\alpha \alpha' + \alpha^{\dagger} \alpha
\alpha'^{\dagger} \alpha')|0 ),
\label{tqc:eq:ent}
\end{equation}
for example, to be implemented
\begin{figure}
\begin{center}
\includegraphics[width=9cm]{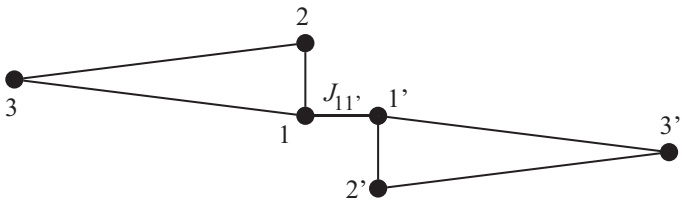}
\end{center}
\caption{Two-qubit system. Physical quantities associated with the
second qubit are denoted with a prime. The coupling strength between
Majorana fermions 1 and 1' is denoted as $J_{11'}$, for example.}
\end{figure}

Now we would like to propose an alternative operation to implement the
state (\ref{tqc:eq:ent}). We first let Majorana fermion $\gamma_1$
of qubit 1 and Majorana fermion $\gamma'_1$ of qubit 2 come closer so that
they interact with each other. The relevant interaction Hamiltonian is
\begin{equation}
iJ_{11'} \gamma_1 \gamma_{1'} = 
iJ_{11'} (\alpha + \alpha^{\dagger})( \alpha' + \alpha'^{\dagger}).
\label{tqc:eq:entstate}
\end{equation}
The interaction strengths are arranged to satisfy 
\begin{equation}
|J_{1 2}|, |J_{1'2'}| \gg |J_{11'}|
\label{tqc:eq:3}
\end{equation}
and
\begin{equation}
|J_{11'}| \gg |J_{12}-J_{1'2'}|.
\label{tqc:eq:4}
\end{equation}
It follows from the condition (\ref{tqc:eq:3}) that the state $\alpha
\alpha'|0 )$ has no time evolution since $J_{11'}$
is negligible compared to $J_{12}+J_{1'2'}$. In contrast,
there is an oscillation between two states $\alpha \alpha'^{\dagger}
\alpha'|0 )$ and $\alpha^{\dagger} \alpha \alpha'|0 )$
since it follows from the condition (\ref{tqc:eq:4}) that $|J_{12}
- J_{1'2'}|$ is negligible compared to $J_{1 1'}$.
Now we are ready to outline how to generate a state like
(\ref{tqc:eq:entstate}).
\begin{enumerate}
\item[STEP 1] We first prepare the state $\alpha \alpha'|0 )$.
\item[STEP 2] Apply $M(\pi/2, \pi/2)$ of Eq.~(\ref{qtc:eq:mgate}) on the
second qubit to generate a state $\alpha \alpha'^{\dagger} \alpha'
|0 )$.
\item[STEP 3] Introduce $J_{11'}$ coupling to transform the state into
$$
\frac{1}{\sqrt{2}}(\alpha \alpha'^{\dagger} \alpha'+\alpha^{\dagger}
 \alpha \alpha')|0).
$$
\item[STEP 4] Apply $M(\pi/2, \pi/2)$ again on the second qubit to obtain the
entangled state
\begin{equation}
\frac{1}{\sqrt{2}}(-\alpha \alpha' + \alpha^{\dagger} \alpha
\alpha'^{\dagger} \alpha')|0 )
\label{tcq:eq:xs}
\end{equation}
as promised. 
\end{enumerate}
We have dropped the operators $\beta$ and $\beta'$ which appear in the
intermediate state.

There is practically no change in the state
(\ref{tcq:eq:xs}) due to the condition (\ref{tqc:eq:3}) once this state
is created.
Qubits 1 and 2 may be widely separated for further stabilization.

\section{Conclusion}

In conclusion, we have proposed new qubit construction in topological
quantum computing, in which Majorana fermions trapped in a two-dimensional
$p$-wave superfluid are employed. A single qubit is constructed out of
three Majorana fermions. An arbitrary one-qubit gate can be
implemented by a combination of the braiding of the vortices (and hence
the Majorana fermions)
and the dynamical phase change. Entangling operation required for
two-qubit gate implementation is shown be realizable in a similar manner.

Introducing a dynamical phase in TQC might seem to be a flaw in an otherwise
perfect quantum computation scheme. It should be noted, however, that a
brading in mathematics, which requires {\it exact} exchange of positions of
Majorana fermions, is never possible to realize physically.
Exchange of positions in reality always involve an imperfection.

\section*{Acknowledgement}

We would like to thank Takeshi Mizushima and Kazunari Machida for 
useful discussions. This work is partially supported by Grant-in-Aid for
Scientific Research (C) from JSPS (Grant No. 19540422).

\end{document}